# An Accessible Formulation for Defining the SI Second Based on Multiple Atomic Transitions

Claudio E. Calosso
Istituto Nazionale di Ricerca Metrologica (INRIM)
Torino, Italy
c.calosso@inrim.it

Nils Nemitz
National Institute of Information
and Communications Technology (NICT)
Tokyo, Japan
Nils.Nemitz@nict.go.jp

*Summary* — The atomic transitions employed in the best of today's optical clocks are a strong foundation for the upcoming redefinition of the SI second. Including multiple transitions in the definition offers increased accuracy, a robust diversity of implementations and would drive continuous performance validation through frequency comparisons. The cost is that such a definition is more complex to articulate and feared to be challenging to implement. We show that it can be made more approachable to intuition, illustration and implementation through formulating this ensemble definition of the SI second in terms of the weighted arithmetic mean of normalized atomic transition frequencies. This definition produces the same results as the presently discussed option up to second order terms of order $10^{-30}$ or below.

## I. Introduction

The redefinition of the second driven by the rush of progress in optical clocks over the last decades [1] occurs in a very different landscape than the preceding one in 1967. Then, the challenge was the transition from an astronomical definition of time to an atomic standard. But cesium was clearly the preferred choice for this standard, and its advantages are evident in the longevity of that definition.

For the redefinition now envisaged for the 29th General Conference on Weights and Measures (CGPM) in 2030 [2], the challenge is how to take advantage of multiple transitions and technologies with comparable performance that all surpass the limits of the cesium standard. The strongest alternative to simply replacing the reference transition (Option 1) is a definition based on multiple transitions [3] chosen from those probed by state-of-the-art clocks (Option 2). Besides the accuracy benefit from averaging over a larger number of different clock implementations, this places co-primary frequency standards based on diverse technologies on an equal footing and promotes competitive development. This diversity also reduces the impact of unanticipated physical effects, such as the discovery of the Black Body Radiation shift of the transition defining the cesium second [4].

The challenge of an ensemble definition is that a complete implementation requires multiple clocks and is often impractical. A complex formulation of the definition further complicates the discussion how such a definition might be put into practice. Here we aim to present such a definition in an accessible form that supports well-informed choices for the future SI second.

## II. A Definition Based on Normalized Frequencies

*The second, symbol s, is the SI unit of time. It is defined by taking the fixed numerical value of the caesium frequency $\nu^\star_{\text{Cs}}$, the unperturbed ground-state hyperfine transition frequency of the caesium 133 atom, to be 9 192 631 770 when expressed in the unit Hz, which is equal to $s^{-1}$.* [5]

We can write this definition of the cesium second as

$$\frac{\nu^\star_{\text{Cs}}}{K_{\text{Cs}}} \triangleq 1 \text{ Hz} . \qquad (1)$$

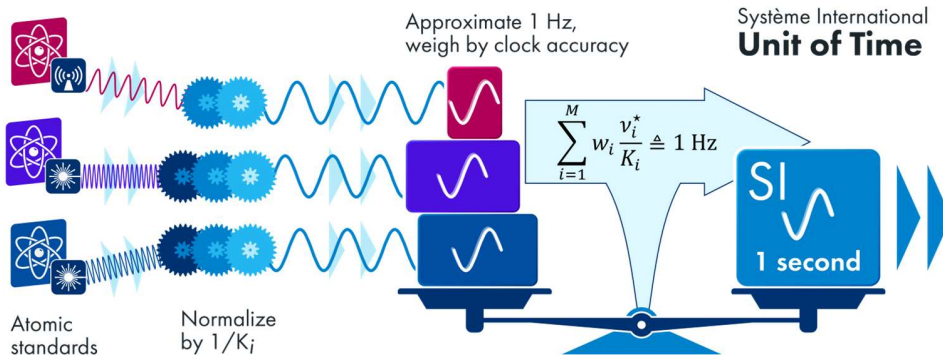

**Figure 1.** Defining the SI second by the weighted mean of normalized frequencies.

From a timekeeper's perspective, the different frequencies of atomic standards are easily converted to a convenient reference frequency. The normalized frequency of each standard already provides an implementation of the SI second with a defined uncertainty. The ensemble improves accuracy and robustness.

Fixing the normalizing constants $K_i$ and weights $w_i$ combines the true atomic transition frequencies $\nu^\star_i$ into a 1 Hz frequency that defines an equally true SI second.

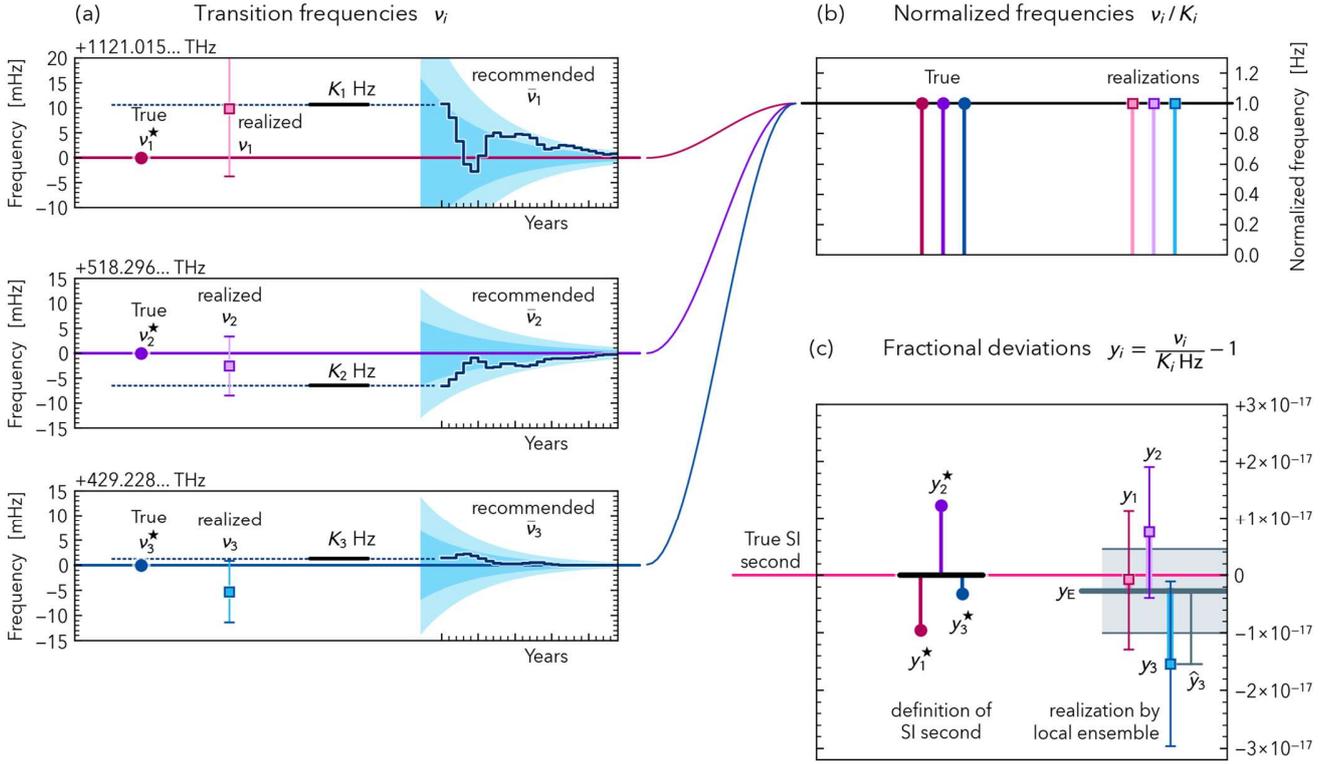

**Figure 2. (a)** In an ensemble definition, the normalizing constants $K_i$ do not define the immediate numerical values of the true transition frequencies $v_i^\star$, they instead reflect the knowledge of clock frequency ratios available at the time of definition and remain fixed at this value. The accuracy of realizing the SI second with a single clock is instead improved by regular publication of recommended frequencies $\bar{v}_i$. Blue 1σ-/2σ-bands show the improvement in their accuracy. Due to the uncertainty of the applied clock, the realization $v_i$ additionally differs from $v_i^\star$. **(b)** The clock frequencies normalized by $K_i$ differ from 1 Hz only within the uncertainties in realizing $v_i$ and assigning $K_i$. **(c)** They are best expressed by fractional deviations $|y_i| < 10^{-15}$. The weighted mean over $y_i^\star$ defines the true SI second, and the weighted mean over $y_i$ realizes this definition with an uncertainty (gray 1σ-band) lower than the individual clocks (error bars). The deviation of the individual realization from the ensemble mean is described by the measure $\hat{y}_i = [y_i - y_E]$.

The defining constant was chosen to have the exact value of $K_{Cs} = 9{,}192{,}631{,}770$ to give the cesium frequency in terms of the previous astronomical definition to the limit of the most accurate measurements. It can be seen to normalize the frequency: Where the maker of the clock is interested in $v_{Cs}^\star$, as timekeepers our focus is the reference frequency that the clock provides independent of its architecture. In the context of the definition, we can consider $f_{\text{ref}} = 1\,\text{Hz} = 1\,\text{s}^{-1}$ a one pulse per second signal that is the foundation of a timescale.

A weighted mean over the output of multiple clocks is a straightforward way to improve this timescale, and by working with normalized reference frequencies we are not restricted to cesium clocks.

$$\sum_{i=1}^{M} w_i \frac{v_i^\star}{K_i} \triangleq 1\,\text{Hz} \qquad (2)$$

then becomes a logical extension of the definition (1) to include multiple atomic transitions, as illustrated in figure 1. Appendix A shows that it is mathematically equivalent to the currently discussed formulation.

### III. DEFINITION AND REALIZATION

In (2), the normalizing constants do not directly define the numerical values of the individual clock transition frequencies, since this would require perfect knowledge of their ratios. That makes it important to internalize the distinctions between the clock transition frequency $v_i^\star$, its realizations $v_i$, and the chosen normalizing constant $K_i$ that is applied to both (figure 2).

Although their numerical values are now unknowable, the clock transition frequencies $v_i^\star$ remain unchanging, perfectly accurate, and 'True'. Choosing any fixed set of normalizing constants $K_i$ and weights $w_i$ combines the true $v_i^\star$ to define an equally true SI second. We will find it *convenient* to choose $w_i$ according to the accuracies of the clocks, and $K_i$ such that $v_i^\star/K_i \approx 1\,\text{Hz}$ according to comparisons of optical clocks with the current cesium standard and with each other. The true $v_i^\star$, its realization $v_i$ by an individual clock, and ($K_i$ Hz) then all agree to within the uncertainties of atomic clocks, and the normalized frequencies are best considered in terms of a fractional deviation

$$y_i = \frac{v_i}{K_i\,\text{Hz}} - 1 \iff v_i = (1 + y_i)\,K_i\,\text{Hz}. \qquad (3)$$

Reference [7] gives a good summary of this common method. Equation (2) can then be expressed as

$$\sum_{i=1}^{M} w_i y_i^\star \triangleq 0 \qquad (4)$$

and locates the SI second by the barycenter of the deviations.

An ensemble of clocks that provide realizations $y_i$ with fractional uncertainties $u_i$, can then realize the ensemble SI second with a deviation

$$y_E = \sum_{i=1}^{M} w_i y_i . \qquad (5)$$

The errors in the realizations of different clock transition frequencies are generally uncorrelated, such that the uncertainty $u_E$ is simply the standard error of the weighted mean:

$$u_E^2 = \sum_{i=1}^{M} w_i^2 u_i^2 \qquad (6)$$

This reduced uncertainty is a key appeal of an ensemble definition: An ensemble of $M$ equally weighted transitions realized with uncertainty $u$ allows for $u_E = u/\sqrt{M}$. The effect of any unexpected correction to one of the frequencies would also be mitigated by the factor $M$.

## IV. MEASURES AND ADJUSTMENTS

While the illustrations are drawn with perfect knowledge of the true SI second, a practical realization of the clock ensemble instead produces results in terms of the frequency ratios

$$\rho_{i,j} = \frac{\nu_i}{\nu_j} = \frac{K_i(1+y_i)}{K_j(1+y_j)} \simeq \frac{K_i}{K_j}\left(1 + [y_i - y_j]\right). \qquad (7)$$

The approximation allows us to identify any ratio of clock frequencies as the nominal ratio $K_i/K_j$ plus a measurable fractional difference

$$y_{i,j} \triangleq [y_i - y_j] = \rho_{i,j}\frac{K_j}{K_i} - 1. \qquad (8)$$

As long as we are working with atomic frequency standards, our choice of $K_j$ makes $|y_j| < 10^{-15}$ and limits the second order terms neglected in (7) to less than $10^{-30}$. The shorthand notation $y_{i,j}$ will reappear through the text.

The frequency of any clock $i$ connected to the ensemble is characterized by the measure

$$\hat{y}_i = [y_i - y_E] = \sum_{j=1}^{M} w_j [y_i - y_j] \qquad (9)$$

of the fractional deviation from the ensemble mean, which is equivalent to the weighted mean of the measures of clock $i$ with each clock of the ensemble.

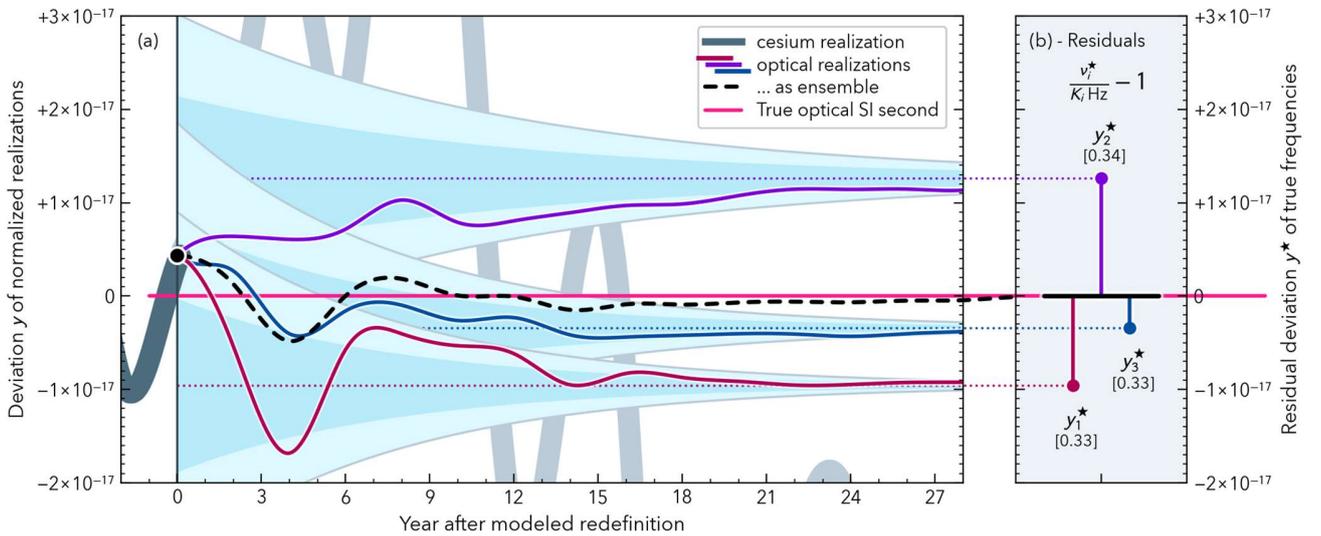

**Figure 3.** **(a)** Illustrating the progress of optical realizations. At the time of redefinition, the normalizing constants $K_i$ are chosen so that the optical realizations of the SI second match the cesium realization (black circle). Even after the redefinition the values of $(K_i \text{ Hz})$ are not equal to the true frequencies $\nu_i^\star$ due to the uncertainties of the optical frequency measurements (blue 1σ-/2σ-bands). While the ensemble mean of the optical realizations (dashed line) steadily approaches the true value, reductions in clock uncertainties eventually reveal residual deviations of the individual realizations $y_i$. **(b)** The realizations are converging towards the normalized true transition frequencies $y_i^\star$. In the ensemble definition, the normalizing constants $K_i \approx \nu_i^\star/\text{Hz}$ make $y_i^\star \approx 0$. The choice of weights (shown in square brackets) then affects the barycenter defining the true SI second only within the bounds of the residuals.
Near-equal weights were assigned for the figures to visualize the definition by the barycenter. To illustrate the continuity of realizations across the redefinition, the model assumes that the values $K_i$ are determined from the clock data that provides realizations, e.g. through contributions to international time. The continuity of the realization will in reality be partially obscured by separate measurement noise contributions in $y_i$ and $K_i$.

In this way, it is straightforward to determine $\hat{y}_i$ with respect to the ensemble, and it is feasible to correct a standard's output for this difference in real-time [6]. The adjusted reference frequency may then replicate the ensemble-realized SI second beyond the limit of the systematic uncertainty of the clock acting as the source oscillator. A clock with sufficient stability can also serve as a hold-over oscillator during any interruption of the full ensemble.

## V. CO-PRIMARY REALIZATIONS

The same correction strategy also enables realizations of the SI second where the full ensemble is not available. As optical clocks improve in accuracy, they eventually resolve the residual deviations $y_i^\star$ of the normalized true transition frequencies from the ensemble mean that defines the SI second (figure 3).

To allow a single clock to provide the best available realization of the SI second, its normalized frequency needs to be corrected for this residual deviation, using a separately determined best estimate of $y_i^\star$. This estimate is the recommended frequency $\bar{y}_i$. To find it we look to the work of the CIPM Working Group on Frequency Standards, which maintains a matrix $\bar{\rho}_{i,j}$ of frequency ratios condensed from all published clock comparisons [8, 10]. Using (8) provides the estimate

$$\bar{y}_i = \sum_{j=1}^{M} w_j \underbrace{[\bar{y}_i - \bar{y}_j]}_{\bar{y}_{i,j}} = \sum_{j=1}^{M} w_j \left( \bar{\rho}_{i,j} \frac{K_j}{K_i} - 1 \right). \quad (10)$$

According to (3), we then find the desired approximation

$$\bar{\nu}_i = (1 + \bar{y}_i) K_i \text{ Hz} = \sum_{j=1}^{M} w_j \, \bar{\rho}_{i,j} \, K_j \text{ Hz} \quad (11)$$

of the true frequency $\nu_i^\star$, to use as the recommended frequency for realizing the SI second with a single standard. The Working Group already provides such recommended frequencies for the best-practices realization of the second by secondary frequency standards. The associate fractional uncertainty $\bar{u}_i$ represents the limit to our collected knowledge of the true transition frequencies rather than any technical properties of the local clock used to realize the SI second. It therefore appears in addition to the statistical and systematic clock uncertainties $u_A$ and $u_B$. In contrast to the full ensemble realization discussed before, $\bar{\nu}_i$ is determined in advance and transmitted in the form of a number, rather than a frequency signal.

Figure 4 illustrates how the same regular updates to the recommended frequencies enable single-standard realizations that converge towards the true duration of the SI second. All single-standard realizations are then handled in this way, and an argument can be made that an ensemble definition simply considers all standards to be secondary representations of the SI second. We would like to offer the complementary view that all "co-primary" standards benefit from their inclusion.

For illustration, we consider an idealized case where all previous information on frequency ratios comes from a remote toy ensemble. This includes one clock per transition, and all have been operated simultaneously until they achieved their uncertainties $u_i$. Replacing the condensed frequency ratios $\bar{\rho}_{i,j}$ in (11) with $\rho_{i,j} = \nu_i/\nu_j$ from the toy ensemble, we find (Appendix B) the uncertainty of each recommended frequency

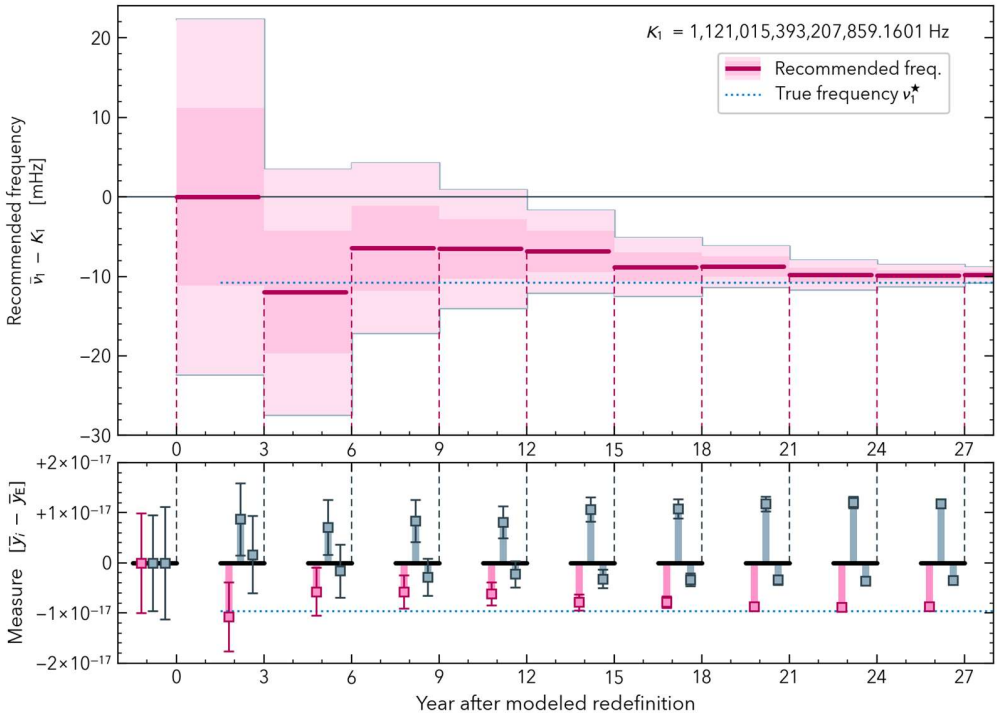

**Figure 4. (top)** Improved determinations of the recommended frequency $\bar{\nu}_1$ (shown as magenta 1σ-/2σ-bands) gradually converge on the true frequency $\nu_1^\star$ (dotted line). This allows single-clock realizations to approach the true value of the SI second without the need to revise the normalization constant $K_1$. **(bottom)** The recommended frequencies represent the expected residual deviation of the normalized frequency from the ensemble mean, and are estimated from the recommended ratios through (10) and (11). Error bars indicate the expected uncertainty relative to the ensemble mean, calculated from a large set of published clock comparisons.

$$\bar{u}_i^2 = (1-w_i)^2 u_i^2 + \sum_{j \neq i} w_j^2 u_j^2 \ . \quad (12)$$

For $w_i = 0$, this frequency standard does not contribute to the ensemble, and the added uncertainty in using it to realize the second is given by $u_i^2 + u_E^2$, just like it would be given by $u_i^2 + u_{Cs}^2$ for a secondary representation of the cesium second. Such *secondary realizations* still benefit from the lower ensemble uncertainty $u_E$.

For $w_i = 1$, we have made this frequency standard the primary representation of the second, since no other frequency standards contribute. Then $u_E = u_i$ (6), and the added uncertainty becomes $\bar{u}_i = 0$, as expected.

The ensemble definition extends the binary choice of primary or secondary realizations into a continuous range characterized by the weight of their contribution. For a *co-primary realization* with $w_i > 0$, both terms of the added uncertainty $\bar{u}_i$ are reduced by the weight $w_i$ assigned to this type of standard. With these reductions, the benefit of the lowered ensemble uncertainty, and the absence of cesium as the dominant source of uncertainty, many future clocks may then operate in the condition $\bar{u}_i^2 \ll u_A^2 + u_B^2$, where they realize the SI second to the limit of their own accuracy. That this depends on the accuracy of $\bar{\rho}_{i,j}$ is a benefit: It creates motivation for each institute operating a highly accurate clock to compare it to other clocks and publish the results.

We can generalize (12) to a partial ensemble $\subset$ of multiple types of standards operating locally. Using correction data from our toy ensemble, this further reduces the added uncertainty according to

$$\bar{u}_\subset^2 = \sum_{j \in \subset} (w_j' - w_j)^2 u_j^2 + \sum_{j \notin \subset} w_j^2 u_j^2 \quad (13)$$

where $w_i'$ is the renormalized weight that the standard $i$ has in the partial ensemble.

For most clocks that provide SI traceable realizations of the second for practical applications, it will be sufficient to simply apply the published recommended frequency $\bar{\nu}_i$ to create a normalized reference signal, and to include the published value of $\bar{u}_i$ in the uncertainty evaluation. The roadmap to the redefinition of the second [2] requires optical frequency ratios to be measured to less then $5 \times 10^{-18}$ uncertainty, such that most or all of the uncertainties $\bar{u}_i$ will fall below this value at the time of the redefinition.

## VI. INTRODUCING THE OPTICAL ENSEMBLE SECOND

Each of the defining constants in (2) has a clear significance. The normalizing constants $K_i$ represent physical properties of the atomic transitions, now investigated to 18 digits of precision.

The weights represent the technical properties of the atomic clocks expected to probe these transitions over the course of the definition, which are difficult to judge or predict to more than two digits of precision. It is the precise normalization performed in the first step that makes the definition of the second insensitive to the informed, but ultimately arbitrary choice of weights. In a definition based on a single transition, this choice is simply hidden behind the self-imposed condition that all $w_i$ must be either 0 or 1.

To redefine the second using an ensemble definition, the first step is to select the $M$ transitions of interest. This choice is driven by scientific and technical concerns, such as the accuracy of realizing the transition frequency now and in the predictable future, and the technical complexity of doing so. It may also include political considerations such as ensuring that the National Metrology Institutes can provide frequency standards according to the local letter of the law.

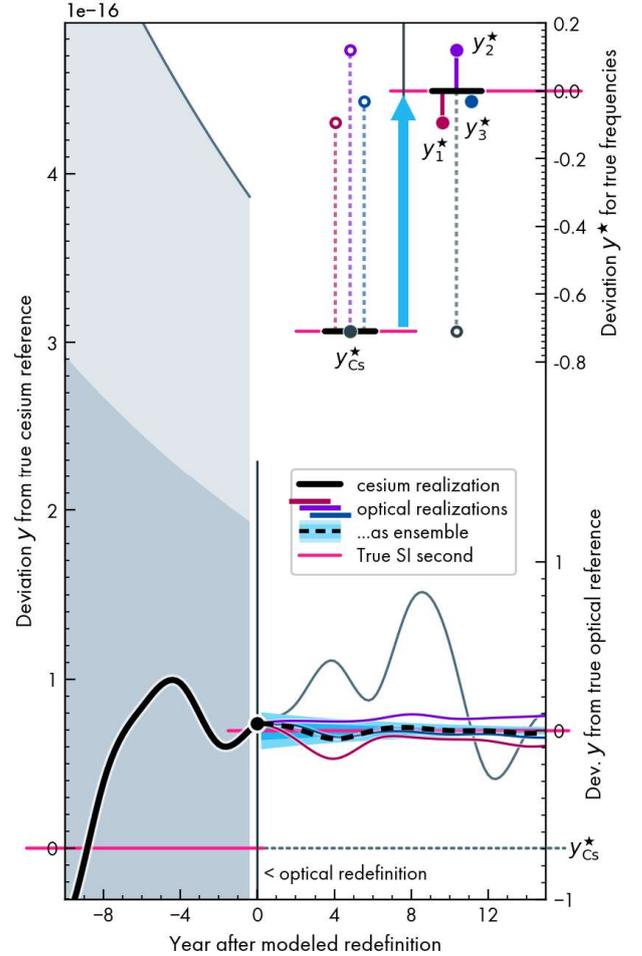

**Figure 5:** Transition from the cesium second to an optical ensemble. The gray 1σ-/2σ-band shows the modeled uncertainty of the cesium realization (black line) of the SI second. The redefinition is prepared by assigning normalizing constants $K_i = \bar{\nu}_i/\text{Hz}$ according to the latest recommended frequencies, to bring the optical realizations into agreement with the cesium realization (black circle). The inaccessible residuals of the normalized true transition frequencies $y_i^\star$ (top inset) nevertheless remain non-zero due to the uncertainty of $\bar{\nu}_i$. The residuals of the optical realization appear in a tight grouping since the dataset accounts for precise optical-to-optical comparisons, so that the errors are strongly correlated.
The redefinition of the second occurs by shifting the weight from the cesium transition (left segment) to the optical transitions (right segment). The unavoidable change in the definition of the true SI second (magenta lines and blue arrow) is bounded by the magnitude of the residuals. After this initial change, the realization of the optical ensemble provides much lower uncertainties (blue 1σ-/2σ-band).
As before, the model emphasizes the continuity of the realizations by neglecting non-common measurement noise of the clocks that perform ratio measurements and those that provide realizations.

The second step is to define normalizing constants $K_i$. Since the true frequencies $v_i^\star$ are inaccessible, the most stringent requirement for continuity is that realizations shall produce the same normalized frequency before and after the redefinition. While this is most critical for the ensemble mean, it is beneficial if the normalized frequencies of different standards agree with each other. A suitable choice is then $K_i = \bar{v}_i/\text{Hz}$, where $\bar{v}_i$ are the recommended frequencies provided by the Working Group on Frequency Standards [8, 10]. Calculated in close analogy to (11), these not only represent frequency measurements relative to the cesium standard but also reflect the more precise frequency ratios found in direct comparisons of optical clocks.

The redefinition is completed with the third step of selecting weights $w_i$ (which sum to one). A guiding principle is to assign weights proportional to the inverse square of the fractional uncertainty contribution, which minimizes the uncertainty of the weighted mean. But since the weights are fixed for the lifetime of the definition, the relevant uncertainties are those of future measurements, and of clocks that may not even have been designed yet. Thus, the assignment remains at least partly arbitrary, and weights might be based on the performance of the best reported implementation of each standard, or an effective value [11] may be determined from the available comparison data [8, 10].

The newly chosen $K_i$ minimize the residual deviations $y_i^\star$, but cannot make their unknown values zero entirely, such that assigning new weights affects the definition of the SI second according to (4), as illustrated in figure 5. Such a change is unavoidable due to the uncertainty

$$u_{\Delta E} = \sqrt{\tilde{u}_{Cs}^2 + \tilde{u}_E^2} \;. \tag{14}$$

of comparing the old and new definition of the true SI seconds. Here $\tilde{u}$ extends the concept of the realization uncertainty $u$ for an individual transition $i$ or an ensemble to describe the combined contributions to the matrix $\bar{\rho}_{i,j}$ [8, 10]. Once enough independent clocks contribute such data, the generalized realization uncertainty $\tilde{u}_i$ of a transition will be smaller than the $u_i$ of the realizations by individual clocks.

Finding $\tilde{u}_i$ from the $u_i$ of the input data requires a careful consideration of the correlations and reproduction of the uncertainty adjustments applied in the determination of $\bar{\rho}_{i,j}$. A more convenient estimate based on the output of this determination is to apply an N-corner-hat method to the fractional uncertainties $\bar{u}_{i,j}$ provided for the ratios $\bar{\rho}_{i,j}$, as also suggested in [11]. Reasonable results are obtained by

$$\tilde{u}_i^2 = \frac{1}{M-1} \sum_{j \neq i} \left( \bar{u}_{i,j}^2 - \sum_{k \neq i,j} \frac{\bar{u}_{k,j}^2}{2(M-2)} \right) \tag{15}$$

which yields $\tilde{u}_{Cs} = 1.9 \times 10^{-16}$ for the main cesium term.

---

The second, symbol s, is the SI unit of time. It is defined by taking the weighted mean of $M$ normalized frequencies $v_i^\star$, each representing an unperturbed atomic transition, to be 1 in the unit Hz, which is equal to $s^{-1}$.

$$\sum_{i=1}^{M} w_i \frac{v_i^\star}{K_i} = 1 \text{ Hz} \qquad \text{with weights} \qquad \sum_{i=1}^{M} w_i = 1$$

The normalizing constants $K_i$ are chosen to represent the transition frequencies and to ensure the continuity with previous definitions. The assigned weights $w_i$ reflect the accuracy of realizing each unperturbed frequency $v_i$. Taken together, the tabulated constants $K_i$ and $w_i$ provide an unambiguous and universal definition of the SI second.

| Transition | | Normalizing constant $K_i$ | Weight $w_i$ |
|---|---|---|---|
| $^{27}$Al+ ion | $^1S_0 \leftrightarrow {^3P_0}$ | 1,121,015,393,207,859.16 | 0.31 |
| neutral $^{171}$Yb | $^1S_0 \leftrightarrow {^3P_0}$ | 518,295,836,590,863.63 | 0.45 |
| neutral $^{87}$Sr | $^1S_0 \leftrightarrow {^3P_0}$ | 429,228,004,229,872.99 | 0.24 |

**Textbox 1: An example definition of the SI second as the weighted mean of multiple normalized atomic transition frequencies.** Numerical values are included for illustration only and have been obtained from data up to 2021 [Margolis 2024]. The redefinition envisioned for 2030 would benefit from additional data, and would include one or more additional digits for the normalizing constants to accurately reproduce the measured clock frequency ratios.

**Table 1:** Uncertainty contributions for selected reference transitions estimated from the evaluation of the 2021 recommended values of standard frequencies, and corresponding example weights before ($w_i$) and after ($w_i'$) the introduction of the ensemble SI second.

| Transition | $^{133}$Cs ($\tilde{u}_{Cs}$) | $^{27}$Al+ ($\tilde{u}_1$) | $^{171}$Yb ($\tilde{u}_2$) | $^{87}$Sr ($\tilde{u}_3$) |
|---|---|---|---|---|
| uncertainty | $1.9 \times 10^{-16}$ | $9.3 \times 10^{-18}$ | $7.7 \times 10^{-18}$ | $1.1 \times 10^{-17}$ |
| weight $w_i$ | 1.00 | 0.00 | 0.00 | 0.00 |
| weight $w_i'$ | 0.00 | 0.31 | 0.45 | 0.24 |

**Table 2:** Uncertainties in the transition frequencies during a revision of the definition in year 16 of the simulated data shown in figure 6. Weights are $w_i$ before the revision, and $w_i'$ afterwards.

| Transition | $\nu_1$ | $\nu_2$ | $\nu_3$ |
|---|---|---|---|
| uncertainty $\tilde{u}_i$ | $1.17 \times 10^{-18}$ | $2.28 \times 10^{-18}$ | $1.39 \times 10^{-18}$ |
| weight $w_i$ | 0.33 | 0.34 | 0.33 |
| weight $w_i'$ | 0.51 | 0.13 | 0.36 |

The final step in the redefinition is to recalculate the uncertainties $\bar{u}_i$ of the recommended frequencies, since they are a function of the weights, as shown in (12).

VII. EXAMPLE DEFINITION OF AN OPTICAL ENSEMBLE SECOND

Textbox 1 illustrates how a definition of the SI second based on a weighted mean of multiple normalized atomic transition frequencies might be presented. The numerical values are based on data collected until 2021 and evaluated by the Working Group on Frequency Standards [10]. The actual redefinition envisioned for 2030 will benefit from newer data with lower uncertainties.

The normalizing constants are taken directly from the tabulated recommended frequency values, since they already express the ratios of optical-to-optical comparisons in addition to the comparisons to the cesium standard. The weights are determined as $w_i \propto \tilde{u}_i^{-2}$, using uncertainties $\tilde{u}_i$ estimated by the N-corner hat method (15). The results after normalization to $\sum w_i = 1$ and truncation to two decimal places are shown in Table 1. For these weights and uncertainties, the unavoidable change of the true SI second during the redefinition is bounded by $u_{\Delta E} = 1.9 \times 10^{-16}$ due to the uncertainty in the realizations of the cesium second by today's fountain clocks.

VIII. DYNAMIC UPDATES

A proposal for a dynamic definition, Option 2b, calls for regular updates. In this case, a quantitative criterion would be set to trigger a revision when and only if it will provide significant improvement of the realization and dissemination [2].

This is an important qualification, since each revision will introduce a new change in the true SI second according to

$$\Delta y_E = y_E' - y_E = \sum_{i \in S} (w_i' - w_i) y_i^\star, \quad (16)$$

where $w_i$ and $w_i'$ are the assigned weights before and after the revision, and we can estimate the bounds of the change by

$$u_{\Delta E}^2 = \sum_{j \in S} (w_i' - w_i)^2 \tilde{u}_i^2. \quad (17)$$

The relevant values for the revision illustrated in figure 6 (and further examined in Appendix C) are listed in Table 2. We calculate $u_{\Delta E} = 5.2 \times 10^{-1}$, and the simulated data results in $\Delta y_E = -8.1 \times 10^{-19}$, largely consistent with $u_{\Delta E}$. The change is much more tightly bounded than during the initial introduction of an optical second, because the larger uncertainty $\tilde{u}_{Cs}$ no longer contributes, and because $|w_i' - w_i| < 1$ once an ensemble has been established.

Since updates to the recommended frequencies are sufficient to exploit the growing knowledge of the transition frequencies, revisions would only occur if one or more types of frequency standards fall so far behind in the improvement of their uncertainty that the assigned weights no longer result in an effective ensemble mean, or if a new standard warrants inclusion in the ensemble. Despite the continued rapid progress of optical clock performance, new clocks with significantly improved performance usually only enter service every few years [1]. It is then unlikely that changing the weights would be required more often than once per decade.

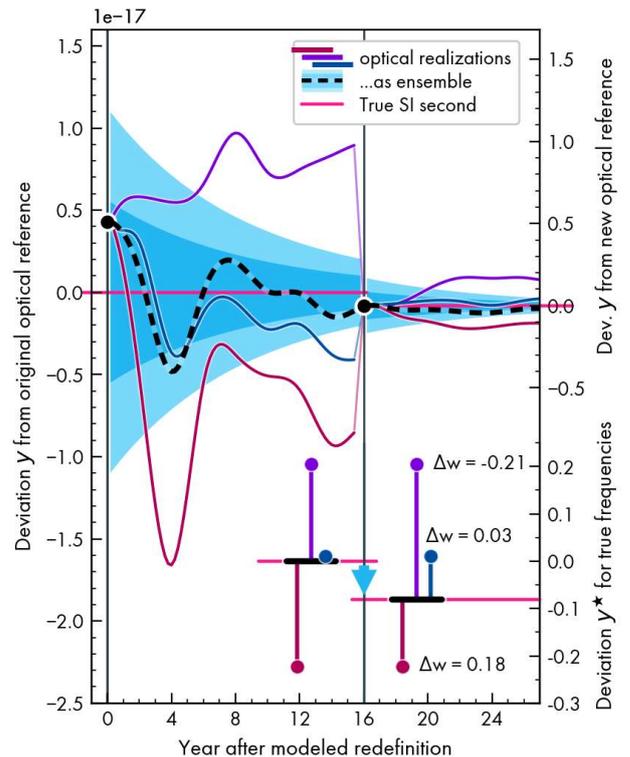

**Figure 6:** A revision of the optical ensemble definition may arise from the need to update the assigned weights. In the simulated data, the accuracy of standard 2 does not improve as quickly as that of the others (figure 3). A recalculation of the weights then transfers approximately 20% of the total weight from standard 2 to standard 1. The resulting change in the ensemble mean of the normalized true frequencies (in the bottom inset) is now bounded by lower uncertainties $\tilde{u}_i$ and smaller weight adjustments $\Delta w$. The conditions that initiate a revision need to be set so that the frequency corresponding to the true SI second converges towards a value within the original confidence band.

A criterion for initiating a revision of the weights might then consider not only that the resulting $u'_E$ is a sufficient improvement over the current $u_E$, but also that $u_{\Delta E}$ is small enough to keep $\Delta y_E$ from accumulating over repeated revisions. When revisions are applied with consideration, then only an outside observer with access to perfect measurements may distinguish the residual changes to the *definition* (figure 6) from the gradual evolution of its *realization* towards this inaccessible true value (figure 3).

## IX. Conclusions

Considering a definition of the SI second based on multiple atomic transitions in terms of the familiar weighted arithmetic mean of normalized frequencies casts it in familiar terms to help understanding and improve intuition. From this vantage point, mathematical complexity no longer needs to hold back the discussion, so that it can focus on the technical challenges and the benefits of such a definition, which supports continued development of frequency standards based on diverse architectures and directly encourages clock comparisons.

Since our formulation differs from the original proposal [3] only by second order terms at a level below $1\times 10^{-30}$, all newly gained insight applies to the Option 2 proposal now investigated by the CCTF Task Force on the Roadmap to the redefinition of the second.

## Acknowledgment

The authors are sincerely grateful to Patrizia Tavella for insightful and stimulating discussions, as well as Jérôme Lodewyck and Tetsuya Ido for their tireless advocacy of the benefits of an ensemble definition. The authors thank the Working Group on Frequency Standards for processing and publishing data crucial to preparing the redefinition.

This work presents scientific findings and perspectives that do not necessarily reflect the stance of the authors' institutions.

## Appendix A: Equivalence of Formulations

The original proposal for a definition of the SI second with a set of optical clock transitions [3] can be written as

$$\prod_{i=1}^{M} (\nu_i^\star)^{w_i} \triangleq N \text{ Hz} , \quad \text{where } N = \prod_{i=1}^{M} K_i^{w_i} \quad \text{(A1)}$$

is chosen for continuity with the previous definition. We divide by $N$ Hz and introduce the deviation $y_i^\star = \nu_i^\star/(K_i \text{ Hz}) - 1$:

$$\prod_{i=1}^{M} \left(\frac{\nu_i^\star}{K_i \text{ Hz}}\right)^{w_i} = \prod_{i=1}^{M} (1 + y_i^\star)^{w_i} \triangleq 1 \quad \text{(A2)}$$

Here $|y_i^\star| < 10^{-15}$, since the $K_i$ are chosen to equal $\nu_i^\star$ to the limits of the most accurate frequency estimates prior to the redefinition. We can then approximate

$$\prod_{i=1}^{M} (1 + y_i^\star)^{w_i} \simeq \prod_{i=1}^{M} (1 + w_i y_i^\star) \simeq 1 + \sum_{i=1}^{M} w_i y_i^\star \quad \text{(A3)}$$

where the neglected second order terms in each expansion are of order $1 \times 10^{-3}$ or below. Reinserting this result in (A2) yields (4) and subsequently (2):

$$\sum_{i=1}^{M} w_i y_i^\star \triangleq 0 \quad \Leftrightarrow \quad \sum_{i=1}^{M} w_i \frac{\nu_i^\star}{K_i} \triangleq 1 \text{ Hz} \quad \text{(A4)}$$

This shows that in any practical application, the formulation based on the weighted arithmetic mean of the normalized frequencies matches the original formulation based on a weighted geometric mean. Although it is valid to consider (A4) a linearization of (A1) as discussed in [11], the well-known frequencies of modern frequency standards limit the neglected second-order terms to entirely insignificant magnitudes. All the work investigating Option 2 by its proponents and by the CCTF Task Force on the 'Roadmap to the redefinition of the second' equally applies to both.

## Appendix B: Uncertainties of Recommended Frequencies

Equation (11) provides a best estimate $\bar{\nu}_i$ of the true transition frequency $\nu_i^\star$ to use as the recommended frequency for realizing the SI second with a single standard. To determine its uncertainty, we use the shorthand notation of (8) and (10) to represent the condensed frequency ratio data of $\bar{\rho}_{i,j}$. From

$$\bar{y}_i = \sum_{j=1}^{M} w_j \, \bar{y}_{i,j} \, , \quad \text{we find}$$

$$\bar{u}_i^2 = \sum_{j=1}^{M} \sum_{k=1}^{M} w_j \, w_k \bar{u}_{(i,j),(i,k)} \quad \text{(B1)}$$

where the matrix element $\bar{u}_{(i,j),(i,k)}$ gives the covariance of the frequency comparisons $\bar{y}_{i,j}$ and $\bar{y}_{i,k}$. The Working Group on Frequency Standards already finds these covariances for the 2021 dataset of clock comparisons [10], where they are applied to the determination of the current recommended frequencies for secondary realizations of the cesium second.

To provide additional insight, the main text imagines the source of $\bar{\rho}_{i,j}$ as an idealized toy ensemble of uncorrelated clocks. If all clocks in this ensemble are simultaneously compared until their uncertainty contributions reach $u_i$ then each term of the measurement covariance matrix is

$$\bar{u}_{(i,j),(i,k)} = \underbrace{u_i^2}_{A} - \underbrace{\delta_{i,j} u_i^2}_{B} - \underbrace{\delta_{i,k} u_i^2}_{C} + \underbrace{\delta_{j,k} u_j^2}_{D} . \quad \text{(B2)}$$

where $\delta_{a,b}$ is the Kronecker delta, which is 1 when $a = b$ and 0 elsewhere. In other words, the variance is 0 when $j = i$ or $k = i$ since we know ab-initio that $\nu_i^\star/\nu_i^\star = 1$, with no uncertainty. Otherwise, it is $u_i^2 + u_j^2$ for the diagonal elements where $j = k$, and finally $u_i^2$ in the other cases, because they share only the contribution from clock $i$.

For the toy ensemble, the sum of the covariance matrix elements in (B1) leads to four terms associated with those of (B2): The term A sums to $u_i^2$, while B and C contribute $w_i u_i^2$ each. The term D yields $\sum_{j=1}^M w_j^2 u_j^2$. By consolidating these results and extracting $w_i^2 u_j^2$ from the D term, we arrive at (13) via

$$\bar{u}_i^2 = (1 - 2w_i + w_i^2) u_i^2 + \sum_{j \neq i} w_j^2 u_j^2 \ . \quad (B3)$$

We can repeat the same procedure in the case where a partial ensemble $\subset$ is available. From the frequency of this partial ensemble

$$\bar{y}_\subset = \sum_{i \in \subset} w_i' \bar{y}_i = \sum_{i \in \subset} \sum_{j=1}^M w_i' w_j \bar{y}_{i,j} \ , \qquad \text{we find}$$

$$\bar{u}_\subset^2 = \sum_{i \in \subset} \sum_{l \in \subset} \sum_{j=1}^M \sum_{k=1}^M w_i' w_l' w_j w_k \bar{u}_{(i,j),(l,k)} \quad (B4)$$

In the same toy model, for uncorrelated clocks, the covariance matrix of the frequency comparisons is now

$$\bar{u}_{(i,j),(l,k)} = \underbrace{\delta_{i,l} u_i^2}_{A} - \underbrace{\delta_{i,k} u_i^2}_{B} - \underbrace{\delta_{j,l} u_j^2}_{C} + \underbrace{\delta_{j,k} u_j^2}_{D} \ . \quad (B5)$$

which in (B4) sums to four terms: $\sum_{i \in \subset} w_i'^2 u_i^2$, $\sum_{i \in \subset} w_i' w_i u_i^2$, $\sum_{l \in \subset} w_l' w_l u_l^2$ and $\sum_{j=1}^M w_j^2 u_j^2$ respectively. By consolidating these results and extracting $\sum_{i \in \subset} w_i^2 u_i^2$ from the D term, we arrive at (14) via

$$\bar{u}_\subset^2 = \sum_{i \in \subset} (w_i'^2 - 2w_i' w_i + w_i^2) u_i^2 + \sum_{i \notin \subset} w_i^2 u_i^2 \ . \quad (B6)$$

## APPENDIX C: GENERALIZING THE REDEFINITION OF THE SECOND

The procedure for the redefinition can be generalized by considering a superset $S$ that includes all relevant transitions. Transitions that do not contribute to the *initial* definition have weights $w_i = 0$, while transitions that do not contribute to the *revised* definition have weights $w_i' = 0$.

### A. Update of the normalizing constants

We begin by updating the normalizing constants to match the calculated recommended frequencies (11). The new $K_i' = (1 + \bar{y}_i) K_i = \bar{v}_i / \text{Hz}$ have two important qualities:

*1)* **They reflect our best knowledge of the frequency ratios according to $\bar{v}_i = \bar{\rho}_{i,j} \bar{v}_j$.** Since the condensed matrix $\bar{\rho}_{i,j}$ is made self-consistent by reducing the original dataset comparing $N$ transition frequencies to just $N-1$ independently adjusted frequency ratios [Margolis 2015], $\bar{\rho}_{i,k} = \bar{\rho}_{i,j} \bar{\rho}_{j,k}$, and thus

$$\bar{v}_i = \sum_{k \in S} w_k \bar{\rho}_{i,k} K_k \text{ Hz} = \sum_{k \in S} w_k \bar{\rho}_{i,j} \bar{\rho}_{j,k} K_k \text{ Hz} = \bar{\rho}_{i,j} \bar{v}_j$$

$$\text{or} \quad \bar{y}_i = \sum_{k \in S} w_k \bar{y}_{i,k} = \sum_{k \in S} w_k (\bar{y}_{i,j} - \bar{y}_{j,k}) = \bar{y}_{i,j} + \bar{y}_j \quad (C1)$$

*2)* **Normalizing by $K_i'$ does not change the SI second, since the sets $K_i'$ and $K_i$ produce the same ensemble mean.** Updating the normalizing constants to $K_i' = (1 + \bar{y}_i) K_i$ adjusts all normalized frequencies to be $y_i' = y_i - \bar{y}_i$. The resulting change in the ensemble mean (5) is

$$\Delta y_E = y_E' - y_E = \sum_{i \in S} w_i \bar{y}_i \ . \quad (C2)$$

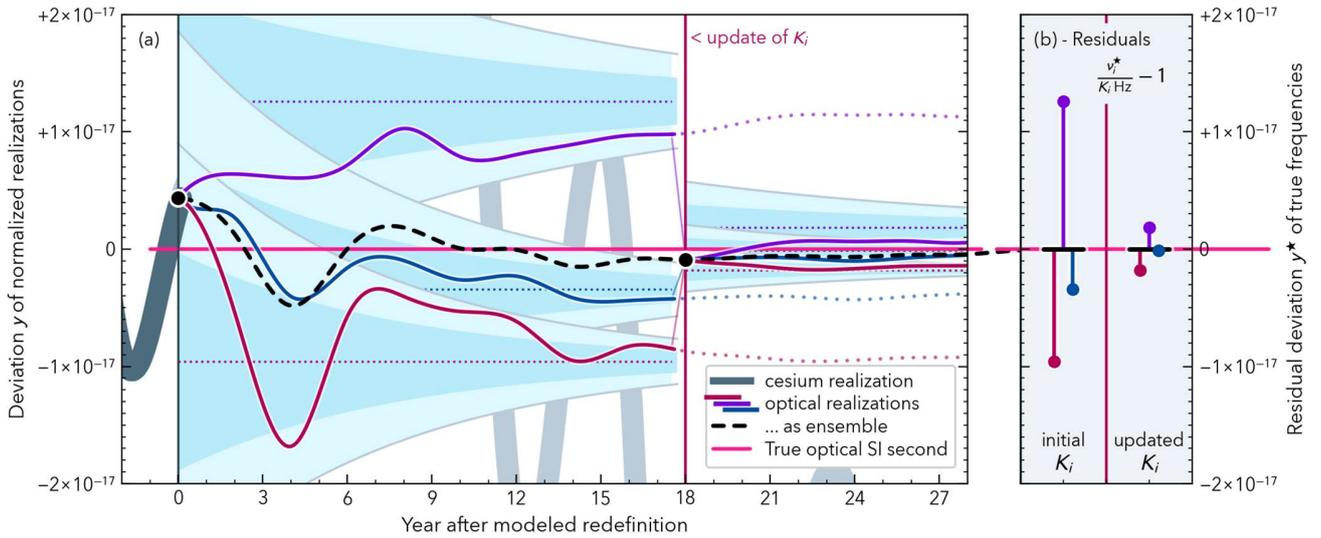

**Figure 7:** A revision of the normalizing constants $K_i$ does not affect the definition or realization of the SI second. **(a)** The revision can be understood to reset the individual realization $y_i$ to the ensemble mean $y_E$ (black circle). To illustrate this, the chosen model neglects non-common noise that is normally present in the frequency ratio measurements used to revise $K_i$ and the realizations $y_i$. If this noise was included, the realizations do not reach complete agreement, but the changes in $y_i$ remain balanced such that the weighted mean is not affected (dashed line). **(b)** The same is true for the residual deviations of the normalized true frequencies $y_i^\star$ (which the realizations converge to, shown by dotted lines). The revision of $K_i$ described in the text minimizes their magnitude while leaving the weighed mean unchanged (magenta line), as well as the true SI second that it defines.

But by inserting $\bar{y}_i$ as written in (10), we find that

$$\Delta y_\mathrm{E} = \sum_{i \in S} \sum_{j \in S} w_i\, w_j\, [\bar{y}_i - \bar{y}_j] = 0 \;, \qquad \text{(C3)}$$

since $w_i\, w_j\, [\bar{y}_i - \bar{y}_j] = -w_j\, w_i\, [\bar{y}_j - \bar{y}_i]$ and all terms cancel.

As illustrated in figure 7, the revision $K_i' = (1 + \bar{y}_i)\, K_i$ aims to make each realization represent the ensemble as accurately as possible by zeroing $y_i$. To better visualize this, the model chosen for figures 4–7 considers $y_i = \bar{y}_i$, where the clock measurement realizing the second are also used to determine the frequency ratios. In this case, normalizing by the revised constants $K_i'$ resets all realizations perfectly to the ensemble mean, which is unlikely to occur in the real-world case. However, this simplification does not affect (C3), which holds for *any* complete set of $y_i$, including for the true $y_i^\star$.

We set $K_i = K_i'$ to complete the update of the normalizing constants. The residual deviations of the normalized true frequencies $y_i^\star$ are now zero to within the uncertainties $\bar{u}_i$ of the recommended frequencies. If we wish to ensure that each $K_i$ remains representative of the most accurate understanding of the atomic transition $i$, this revision could be repeated regularly, since it leaves the SI second unchanged (C3).

There is however no need to do so, because the same information can be conveyed by publishing recommended frequencies as in equation (11) and figure 4. Updating only the normalizing constants $K_i$ expresses the same SI second in terms of different defining constants.

*B. Update of the weights*

The actual redefinition occurs in the following step, which involves the revision of the weights. Any modification will affect the duration of the SI second. However, the adjustment of $y_i^\star$ has minimized the resulting change

$$\Delta y_\mathrm{E} = y_E' - y_E = \sum_{i \in S} (w_i' - w_i)\, y_i^\star \;, \qquad \text{(C5)}$$

where $\Delta y_\mathrm{E}$ is zero within uncertainty $u_{\Delta E}$ estimated by (18).

An example for a change of weights after the initial introduction of the optical transitions is given in figure 6. It illustrates a revision of the weights in response to clock uncertainties that progress at unequal rates (see figure 3), using simulated data. Table 2 shows the modeled uncertainties and weights. We calculate $u_{\Delta \mathrm{E}} = 5.2 \times 10^{-1}$. In reality, $y_i^\star$ and $\Delta y_\mathrm{E}$ are inaccessible. For the simulation, we find a change in the duration of the modeled SI second according to $\Delta y_\mathrm{E} = -8.1 \times 10^{-19}$.

The *original* SI second was realized with the uncertainty $u_\mathrm{E} = 9.8 \times 10^{-19}$. Following the redefinition, the realization still implements the same *original* SI second with $u = \sqrt{u_\mathrm{E}'^2 + u_{\Delta E}^2} \approx 9.8 \times 10^{-19}$. If $u \lesssim u_\mathrm{E}$, the redefinition can be considered to separate $u_\mathrm{E}$ into a reduced $u_\mathrm{E}'$ and a constant $\Delta y_\mathrm{E}$ that is a "frozen" sample drawn from the uncertainty $u_{\Delta E}$. This creates a tighter confidence band without introducing a shift that accumulates over repeated redefinitions.

After the weights have been revised, the final step is again to recalculate the uncertainties of the recommended frequencies according to (B1).